\documentclass[
conference,
10pts
]{IEEEtran}
\IEEEoverridecommandlockouts

\usepackage{amsmath,amssymb,amsfonts}
\usepackage{algorithmic}
\usepackage{graphicx}
\usepackage{textcomp}
\usepackage{xcolor}
\usepackage{bm}
\usepackage{upgreek}
\usepackage{svg}  
\usepackage{url}

\usepackage{balance}
\usepackage{cite}
\def\BibTeX{{\rm B\kern-.05em{\sc i\kern-.025em b}\kern-.08em
    T\kern-.1667em\lower.7ex\hbox{E}\kern-.125emX}}
\begin{document}

\title{Kinetic inductance and voltage response dependence on temperature: Asymmetric dc SQUID case study}

\author{
\IEEEauthorblockN{1\textsuperscript{st} M. A. Gal\'i Labarias}
\IEEEauthorblockA{ \textit{CSIRO Manufacturing}\\
 Lindfield, NSW, Australia. \\
marc.galilabarias@csiro.au}
\and
\IEEEauthorblockN{2\textsuperscript{nd} O. A. Nieves}
\IEEEauthorblockA{ \textit{CSIRO Manufacturing}\\
 Lindfield, NSW, Australia. \\
oscar.nieves@csiro.au}
\and
\IEEEauthorblockN{3\textsuperscript{rd} S. T. Keenan}
\IEEEauthorblockA{ \textit{CSIRO Manufacturing}\\
 Lindfield, NSW, Australia. \\
 shane.ocianain@gmail.com
}
\and
\IEEEauthorblockN{4\textsuperscript{th} E. E. Mitchell}
\IEEEauthorblockA{ \textit{CSIRO Manufacturing}\\
 Lindfield, NSW, Australia. \\
 emma.mitchell@csiro.au}
}

\maketitle

\begin{abstract}
Inductance plays a crucial role in the design and optimization of superconducting quantum interference devices (SQUIDs) for quantum sensing applications, since it dictates the sensitivity and coupling ratio with other circuit elements. In high-temperature superconductors the kinetic inductance, which depends on both geometry and temperature, becomes a dominant part of the device's total self-inductance, since their London penetration depth is considerably larger compared to low-temperature superconductors. In this work, we use an asymmetric SQUID to investigate the kinetic self-inductance ratio and voltage modulation depth at different operating temperatures, device geometries and bias currents.
We first validate our approach by comparing our modelled data with experimental measurements. Then, through numerical simulations, we show: (i) kinetic inductance dominates for thin superconducting films, while for thicker films the inductance is less sensitive to temperature changes; (ii) the voltage modulation depth decreases exponentially with the total inductance independent of the asymmetry ratio; (iii) narrower superconducting tracks lead to a broader temperature operation range, $\Delta T \sim 30 K$, while wider tracks operate in a smaller temperature range, $\Delta T \sim 10 K$, but are more sensitive to temperature changes; and (iv) the device performance versus temperature strongly depends on the bias current used.
\end{abstract}


%
%
%
%
%

\section{Introduction}

Superconducting Quantum Interference Devices (SQUID) have been studied for many years for their high magnetic field sensitivity~\cite{Tesche1977, Koelle1999, Kornev2009} and applications~\cite{Foley1999a, Keenan2010, Leslie2003}.
The voltage-to-magnetic field response is greatly dependent on the inductance of the superconducting elements that constitute the SQUID, but also on the operating temperature.
Therefore, a better understanding of these parameters is crucial for designing an optimal SQUID. 
For instance, at low-temperatures SQUID technologies have been used in thermometry~\cite{Lusher2001,Beyer2007,Beyer2013} for mK measurements, and also Johnson noise based thermometers~\cite{Qu2019}.
Thermometry applications have also been used with high-temperature superconductors (HTS) as devices with a broader applicability range~\cite{Hao1999, Peden1999, Hao2001}.
Other applications of HTS include mineral exploration~\cite{Foley1999a, Leslie2003, Hato2013, Stolz2021}, medical imaging~\cite{Schneiderman2019} or gradiometry~\cite{Hatsukade2011, Stolz2020}.

Proper understanding and characterisation of the inductance and temperature dependence is fundamental for all these applications, since the SQUID inductance determines the coupling with amplifier or pick-up loop, but also affects the device flux noise, thus its performance.
This is even more important for HTS, since due to larger London penetration depths, the kinetic inductance can be the dominant term of the total SQUID self-inductance.
Experimental measurement of the inductance and transfer function of YBCO SQUIDs have been previously done for certain SQUID geometries and Josephson junction (JJ) technologies, e.g. step-edge JJs~\cite{Mitchell2002, Mitchell2003} or nano-slit JJs~\cite{Li2019}.
Investigation into the kinetic inductance contribution and SQUID optimization of YBCO have been reported~\cite{Ruffieux2020}.
Also, a method to estimate the temperature-dependent London penetration depth~\cite{Keenan2021}.

In this work, we investigate the kinetic inductance contribution to the total inductance and its dependence on film thickness and operating temperature. 
Then, for different geometries and bias currents, we analyse the effect of changing the operating temperature on the device performance.
To do so, we introduce a mathematical model for an asymmetric SQUID and validate it by comparing our simulations with experimental data.

\section{Mathematical framework} \label{sec:Model}

In this Section we present a mathematical model for computing the voltage response of a geometrical asymmetric SQUID (see Fig.~\ref{fig:hairpinSQUID_diag}(a)). 
This model assumes homogeneous current densities across superconducting tracks, which is valid when the widths of the superconductive tracks $w$ are smaller than twice the Pearl penetration depth \cite{Pearl1964} $\Lambda = \lambda_L^2 / d$, where $d$ is the film thickness and $\lambda_L$ the London penetration depth \cite{London1935}. We also assume short Josephson junctions where the RSJ model can be used.

\subsection{Asymmetric SQUID modelling}
\begin{figure}[h!]
    \centering
    \includegraphics[width=0.40\textwidth]{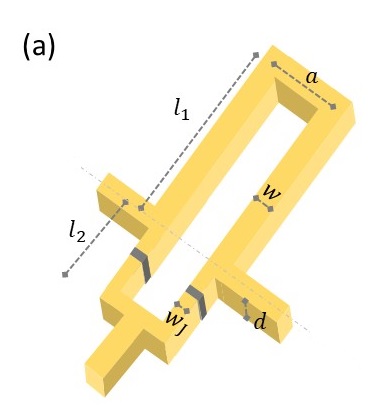}\\
    \includegraphics[width=0.40\textwidth]{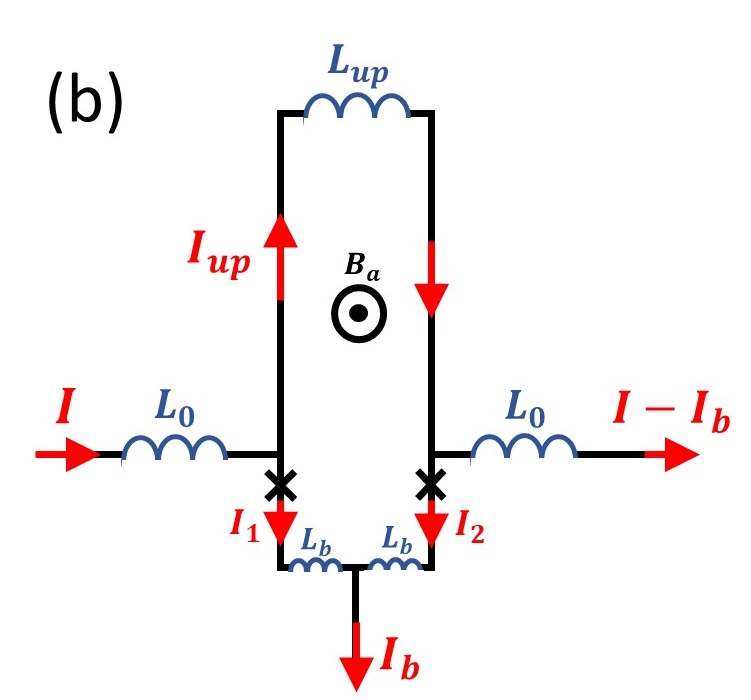}
    \caption{(a) Representation of the asymmetric dc-SQUID under study and its characteristic lengths. (b) Equivalent lumped-element model of the asymmetric dc-SQUID. Equivalent inductors are represented in blue; in red the currents at each section of the SQUID are depicted; the positive direction chosen for the magnetic field is expressed with the traditional notation; finally, the Josephson junctions are represented by the black crosses.}
    \label{fig:hairpinSQUID_diag}
\end{figure}

In this Section we introduce the main equations that model an asymmetric dc-SQUID as shown in Fig.~\ref{fig:hairpinSQUID_diag}.
Similar to the model introduced in \cite{Gali2022a}, we define the total current flowing through each junction as

{}

\begin{equation}
    I_k = I_c\sin ( \varphi_k) + I_{n_k} + \frac{\Phi_0}{2\pi R_n}\frac{d \varphi_k}{dt} \; ,\label{eq:Joseph}
\end{equation}
where $I_k$ is the total current going through the $k^{\text{th}}$ junction, $I_c$ is the critical current and $\varphi_k$ the gauge-invariant phase difference across the $k^{th}$ junction. $I_{n_k}$ is the current created due to thermal noise, $R_n$ is the junction normal resistance and $\Phi_0$ is the flux quantum. 

In previous works \cite{Tesche1977, Gali2023a}, the effects of non-identical Josephson junctions, i.e. $R_{n,1} \neq R_{n,2}$ and $I_{c,1} \neq I_{c,2} $, have been studied. 
In this work, we assume identical junctions, i.e. $R_{n,1} = R_{n,2}=R_n$ and $I_{c,1} = I_{c,2}=I_c$, and focus on the effect of inductance and temperature on the device performance.

Using Kirchhoff's law, we can express the currents $I_1$ and $I_2$ in terms of the input currents $I$ and $I_b$, and the current at the upper part of the loop, $I_{up}$.
\begin{subequations}
\begin{align}
     I_1  &= I - I_{up} \; ,  \\ 
     I_2  &= I_b - I + I_{up}  \; .
    \label{eq:kirch}
\end{align}
\end{subequations}

Using the second Ginzburg-Landau equation, we can express the gauge-invariant phase differences as
\begin{align}
     \frac{\Phi_0 }{2\pi} \left(\varphi_2 - \varphi_1 \right) = \Phi_a - L_s I_{up} + L_1 \left(2I - I_b \right), \label{eq:phase-flux}
\end{align}{}
where $L_s=L_{up} +2 L_b$ is the total SQUID self-inductance, and $L_1 =  L_0 + L_b $, where $L_b$ is the inductance of the bottom part of the asymmetric SQUID. Both $L_s$ and $L_b$ contain kinetic and geometric inductance contributions, and $L_0$ is the inductance of the horizontal bias leads (see Fig.~\ref{fig:hairpinSQUID_diag}(b)). 
For thin superconducting tracks where the supercurrent density can be assumed homogeneous across the track, the kinetic inductance is defined as $L_k(x)=\frac{\mu_0 \lambda_L^2}{d \, w}x$ where $\mu_0$ is the material permeability, $\lambda_L$ the London penetration depth, $d $ and $w$ the film thickness and track width, and $x$ is the length element \cite{Keenan2021, Gali2022a}.
For the SQUID under study the kinetic part of $L_s$ and $L_b$ are 
\begin{subequations}\label{eq:L_k}
\begin{align}
L_{s}^{k} &=\frac{\mu_0 \lambda_L^2}{d} \left( \frac{2l_1 + a}{w} + \frac{2l_2 + a}{w_J} \right) \; , \label{eq:L_ka} \\
L_{b}^{k} &= \frac{\mu_0 \lambda_L^2}{d \, w_J} \left( l_2 + \frac{ a}{2} \right) \; .    
\end{align}
\end{subequations}

Here, we assume that the two horizontal bias leads are identical, so their inductances $L_0$ are the same. 

Using Eqs.~\eqref{eq:Joseph}-\eqref{eq:phase-flux}, we obtain a set of two coupled differential equations that describe the dynamics of the SQUID

\begin{subequations}
\begin{align}
    \dot{\varphi}_1 &=  -\sin(\varphi_1) - i_{n_1} + i + f(\varphi_1, \varphi_2) \; , \\ 
    \dot{\varphi}_2 &= -\sin(\varphi_2) - i_{n_2} + 2i_b - i  - f(\varphi_1, \varphi_2) \; ,
\end{align}
     \label{eq:final_DE}
\end{subequations}
where $i=I/I_c$ and $i_b=I_b/(2I_c)$ are the normalised input currents, $f(\varphi_1, \varphi_2)= \left( \varphi_2 -\varphi_1 - 2\pi\phi_{nf}  \right)/(\pi \beta_L)$ with $\phi_{nf} = \phi_a + L_1 \left(2I - I_b \right)/\Phi_0$ is a time-independent term, $\beta_L=2 I_c L_s/\Phi_0$ is the screening parameter and
$\dot{\varphi_k}=d \varphi_k / d\tau$ indicates the time-derivative where $\tau$ is the normalised time defined as $\tau=2\pi R_n I_c t / \Phi_0$.
The normalized thermal noise current $i_{n_k}=I_{n_k}/I_{c}$ is generated at each time-step using a random number generator which follows a Gaussian distribution with zero mean $\langle i_{n_k}\rangle=0$, and mean-square-deviation $\langle i_{n,k}^2\rangle=\sigma(i_{n_k})^2=2\Gamma/\Delta \tau$, where $\Gamma=2\pi k_B T/(I_{c}\Phi_0)$ is the thermal noise strength with $k_B$ the Boltzmann constant and $T$ the operating temperature, and $\Delta\tau$ is the time-step used in the numerical simulations (see \cite{Tesche1977, Voss1981, Gali2022a} for details).
\subsection{Penetration Depth and Critical Current Temperature Dependence}
In this paper we use the well-known empirical formula~\cite{Buckingham1956, Lee1993} to determine the values of the London penetration depth at different temperatures, 
\begin{equation}
    \lambda_L(T) = \frac{\lambda_L(0)}{\sqrt{1 - \left( T/T_c \right)^2}} \; ,
    \label{eq:lam}
\end{equation}{}
where $T_c$ is the critical temperature of the superconducting material. We use measured $T_c$ and experimental data of YBCO films at $T=77 \,$K to estimate the penetration depth at zero temperature, namely $\lambda_L (0)$. We then use this value and Eq.~\eqref{eq:lam} to calculate $\lambda_L(T)$ at different operating temperatures.

Since the critical current $I_c$ and junction resistance $R_n$ also depend on the operating temperature, we use the following empirical formulas which are fitted to experimental data in~\cite{Beck1995, Gross1997} as well as our measured data for $I_c$ and $R_n$ at $T=77 \,$K:
\begin{equation}
    J_c(T) = \begin{cases}
        2.079\times 10^5\left(0.9 - T/T_c\right) & T<75\\
        4.145\times 10^5\left(1 - T/T_c\right)^{2} &  T\geq 75
    \end{cases},
\end{equation}
where $J_c$ has units of A$/$cm$^2$, and for $I_c(T)$ we have
\begin{equation}\label{eq:Ic_temp}
    I_c(T) = J_c(T)\cdot w_J \cdot d \; ,
\end{equation}
where $w_J$ is the junction width and $d$ the film thickness. The normal resistance $R_n$ is computed from the phenomenological relation $I_c R_n \propto J_c^p$ with $p=0.5$~\cite{Gross1997, Mitchell2010, Mitchell2011b}.

\section{Comparison with Experimental Measurements} \label{sec:Val}
In this section we compare our model with experimental measurements.
First, we verify that our method to calculate inductances for thin structures agrees with experimental measurements.
Next, we test our asymmetric dc-SQUID model by comparing simulated and experimental data of the voltage response depending on the current $I$ and also the \textit{I-V} characteristics of the device.

\subsection{Inductance} \label{ssec:induc}
To measure the inductance of these asymmetric SQUIDs (Fig.~\ref{fig:hairpinSQUID_diag}(b)), the current $I$ is increased by a ramp current $\Delta I$ until a multiple of a flux quantum $n \, \Phi_0$ is measured, with $n$ an integer. 
This same method was used previously~\cite{Ruffieux2020, Keenan2021}. 
Then, the measured inductance is obtained from the following expression:

\begin{equation}
    L_{exp} = \frac{n \cdot \Phi_0}{\Delta I} \; . \label{eq:ExpInd}
\end{equation}

Following the same method and adding a ramp signal $\Delta I$ to Eq.~\eqref{eq:phase-flux}, we obtain
\begin{equation}
    \left( L_s - 2L_1 \right) \Delta I = n \, \Phi_0 \; .
\end{equation}
Therefore, the experimentally measured inductance ($L_{exp}$) corresponds to
\begin{equation}
    L_{exp} = L_{up} - 2 \, L_0 \; ,
    \label{eq:Lexp}
\end{equation}
where $L_{up}$ is the inductance of the upper part of the asymmetric SQUID (Fig.~\ref{fig:hairpinSQUID_diag}(b)).

\begin{figure}[h!]
    \centering
    \includegraphics[width=0.43\textwidth]{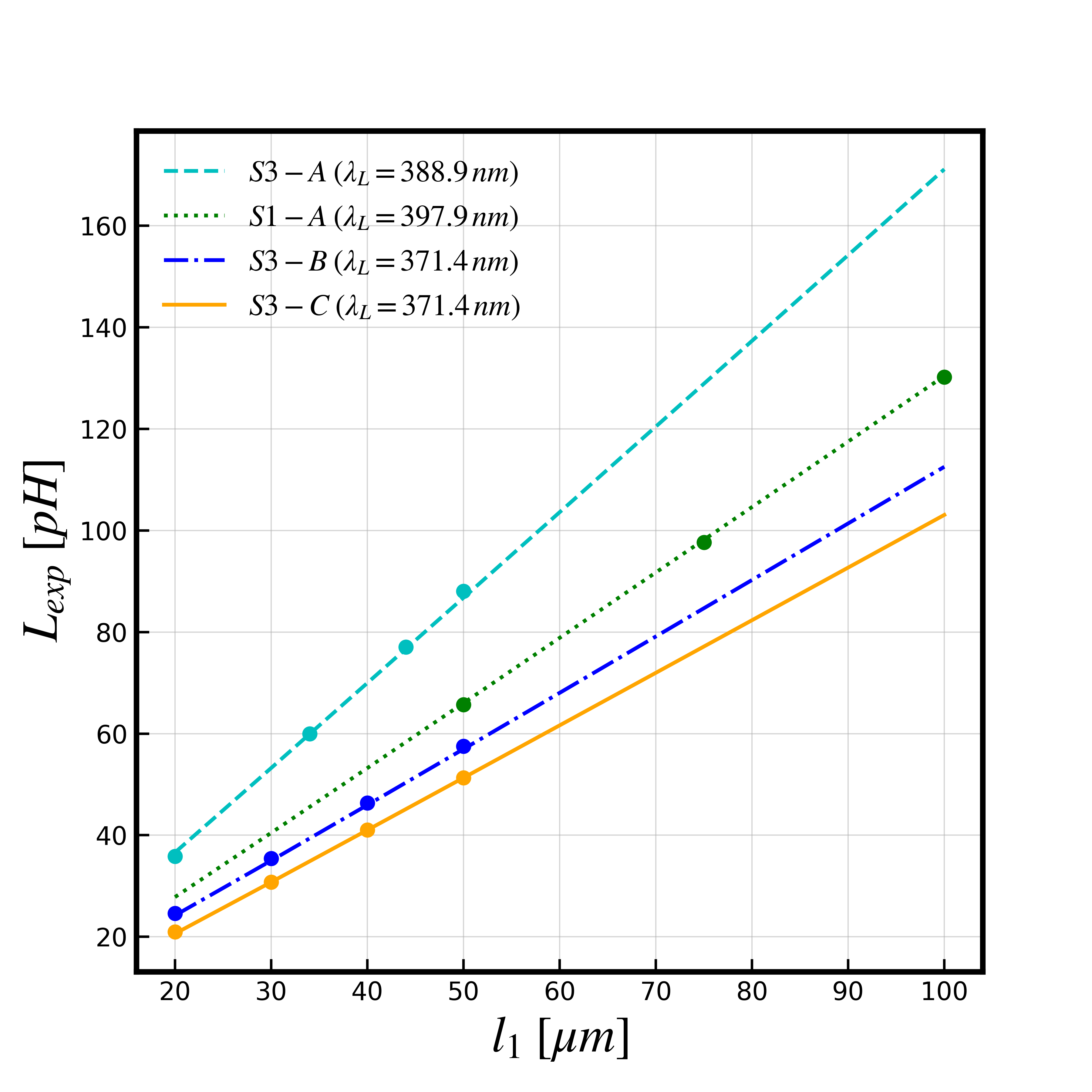}
    \caption{$L_{exp}$ versus the SQUID upper-loop length $l_1$. Experimentally measured data is depicted by dots~\cite{Keenan2021} while lines show $L_{exp}$ values obtained using Eq.~\eqref{eq:Lexp}.
    }
    \label{fig:Lexp}
\end{figure}

Figure~\ref{fig:Lexp} shows the comparison between the calculated and measured inductances for different asymmetric SQUID designs and upper loop lengths $l_1$ (the experimental measurements of these devices have been previously reported in Fig.~2 of \cite{Keenan2021}).
The devices $S3-A,\, B$ and $C$ are from the same batch with $d=113\,$nm, but with different geometry designs, while $S1-A$ has the same design as $S3-A$ but is from a different batch with $d=220\,$nm.
\begin{table}[]
    \centering
    \begin{tabular}{|c|c|c|}
    \hline
         Design & $w$ [$\upmu$m ]  & $a$  [$\upmu$m ]  \\ 
         \hline \hline
         A & 4 & 8 \\
         B & 8 & 12 \\
         C & 8 & 10 \\
    \hline
    \end{tabular}
    \vspace{0.2cm}
    \caption{SQUID designs used in Fig.~\ref{fig:Lexp}. Note that in \cite{Keenan2021} the designs are defined with the parameter $s$ which refers to the inner loop width, thus the correspondence with our notation is $a = w + s$.}
    \label{tab:designs}
\end{table}
The geometric parts of these inductances have been calculated using the expressions for rectangular bar conductors introduced elsewhere~\cite{Hoer1965}.
These results demonstrates that our model and method to calculate $L_{exp}$ is appropriate for this kind of device.

\subsection{Voltage response and $I$-$V$ characteristics}

\begin{figure}[h!]
    \centering
    \includegraphics[width=0.42\textwidth]{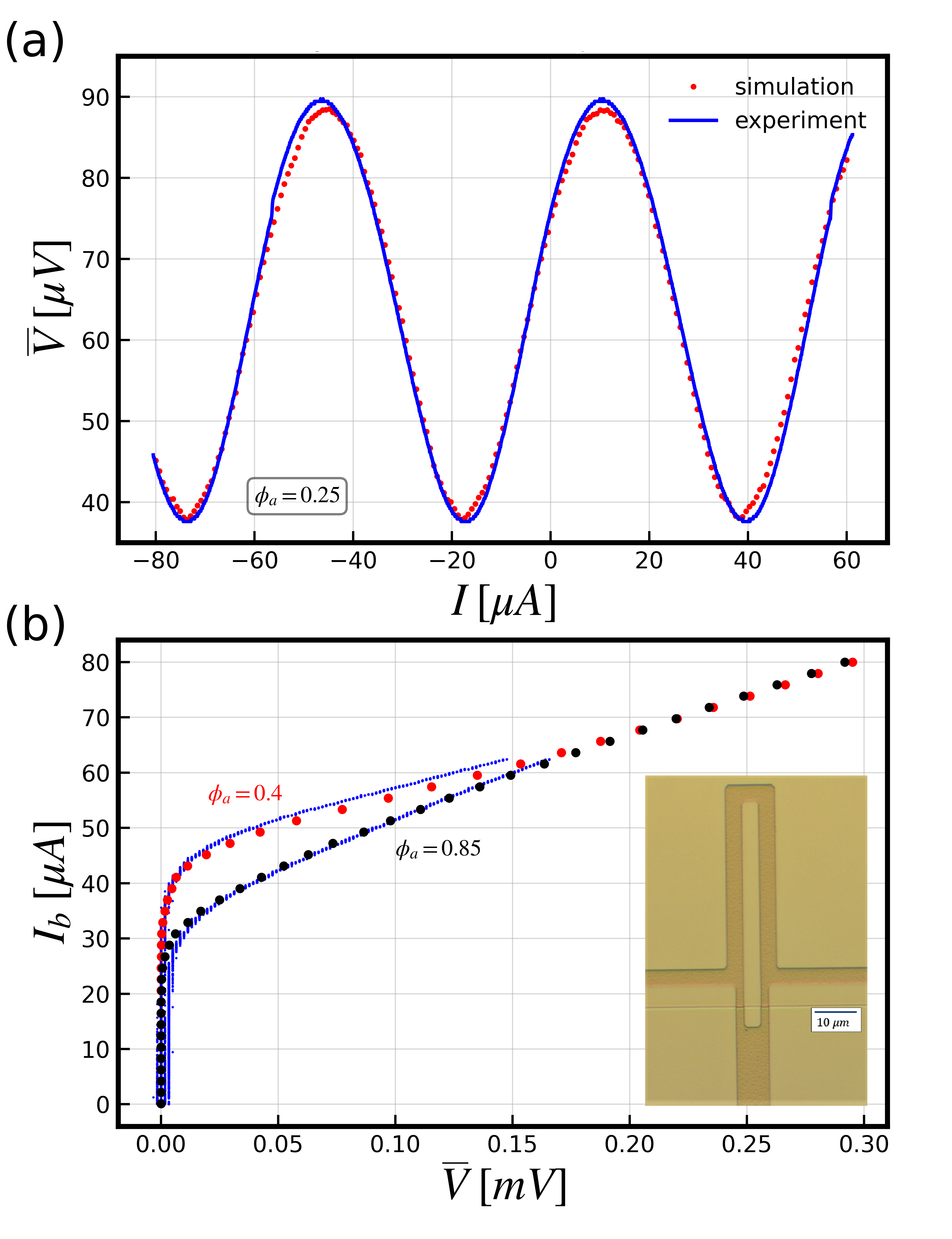}
    \caption{Comparing experimental (blue) and simulated (dots) data of an asymmetric SQUID with $I_c=29.75 \, \upmu $A and $R_n=10.5 \, \Omega$ at $T=77 \, $K. (a) voltage response depending on the input current $I$ with a fixed bias current of $I_b=49 \, \upmu $A and $\phi_a=0.25$. (b) $I$-$V$ characteristics at different applied magnetic fluxes $\phi_a=0.4$ (red) and $\phi_a=0.85$ (black).}
    \label{fig:IV_response}
\end{figure}

The asymmetric dc-SQUID that was used for this experiment is shown in the inset of Fig.~\ref{fig:IV_response}(b), and it has the following dimensions: $a=4\, \upmu $m, $l_1=21.2 \, \upmu $m, $l_2=10 \upmu $m, upper-loop track width of $w=4\, \upmu $m, the junction width is $w_J=2 \, \upmu $m and the film thickness is $d=113 \, $nm. 
From the inset in Fig.~\ref{fig:IV_response}(b) we can see that the geometry of the bottom part of the SQUID is different from our model diagram (Fig.~\ref{fig:hairpinSQUID_diag}).
Here the bottom bias lead is relatively wide which produces flux focusing due to the Meissner currents, as seen previously in wide YBCO structures~\cite{Muller2021}. In order to account for this, we need to introduce an effective area which creates the equivalent total flux, i.e. we have to estimate an effective $l_2$. 

For our simulations we need to determine the $I_c$, $R_n$ and $l_{2,\text{eff}}$ which best fits with the experimental voltage response and $I$-$V$ characteristics. For these simulations $l_{2,\text{eff}}=20\, \upmu $m, $I_c=29.5 \, \upmu $A, $R_n=10.2 \, \Omega$ and $T=77 \, $K.
Figure \ref{fig:IV_response}(a) shows the time-averaged voltage response $\overline{V}$ versus the ramp-current $I$ of experimental (blue line) and simulated (red dots) data. 
Figure \ref{fig:IV_response}(b) depicts the $I$-$V$ characteristics of experimental (blue) and simulated data (circles) at two different applied magnetic fields: $\phi_a=0.4$ (red) and $\phi_a=0.85$ (black).
Our results show a remarkable agreement between our simulation and the experimental measurements.

The experimental data shown in Fig.~\ref{fig:IV_response} has been obtained using Magicon SEL-1 electronics.

\section{Modelling: Results and Discussion} \label{sec:Disc}

In this section we discuss the kinetic self-inductance depending on the film thickness and operating temperature. Then, we investigate the voltage response and voltage modulation depth for different $L_s$, SQUID asymmetric ratios $l_1/l_2$, operating temperatures and bias currents.
The simulation parameters used in this paper are $\Delta \tau =0.1$, $N_t=4\times 10^5$ (total time-steps used) and the device geometry is defined by $w_J=w$, $a=4 \, \upmu $m and $l_2=10 \, \upmu $m and the film critical temperature used is $T_c=86.2 \, $K, which is a common $T_c$ for YBCO thin films. 
At different temperatures $I_c, \; R_n$ and $\lambda_L$ are determined using Eq.~\eqref{eq:lam}. For example, at $T=77\, $K, $\lambda_L= 392.4 \, $nm, $I_c=20 \, \upmu $A and $R_n=10 \, \Omega$.
As shown in Fig.~\ref{fig:IV_response}(a), modulating $I$ creates different induced fluxes, thus without loss of generality we can fix $I$ to a certain value, here we use $I=I_b$, and then only vary the applied magnetic field $\Phi_a$.
Unless specified otherwise, $w=2\, \upmu $m,  $d= 200 \, $nm, $T=77 \, $K and $i_b=I_b/(2I_c)=0.75$.

\subsection{Kinetic self-inductance dependence on geometry and temperature}
\begin{figure}[!h]
    \centering
    \includegraphics[width=0.5\textwidth]{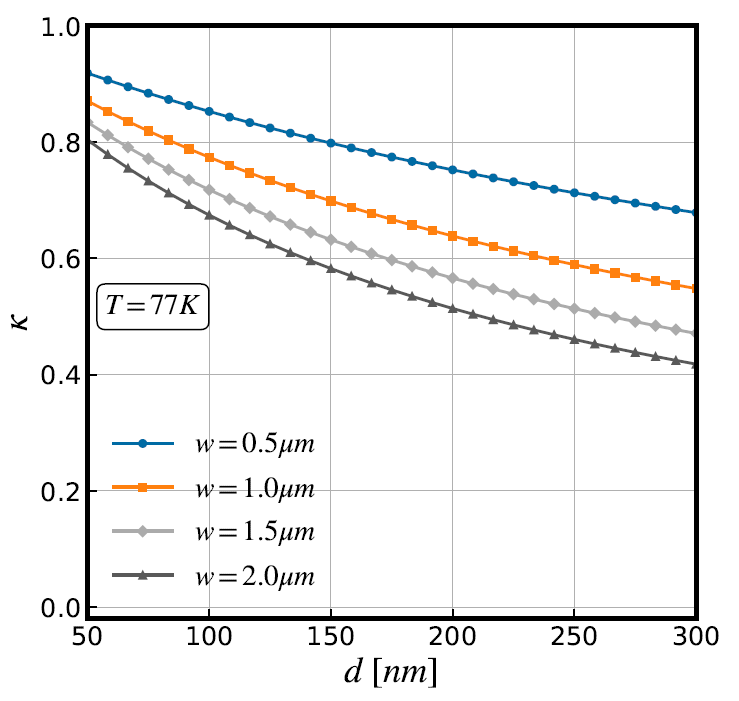}
    \caption{Normalized contribution of the kinetic inductance as a function of film thickness, for devices with different upper track widths $w$ and fixed $l_1=50 \, \upmu m$ at $77 \, $K.}
    \label{fig:kappa_vs_d}
\end{figure}

\begin{figure}[!h]
    \centering
    \includegraphics[width=0.5\textwidth]{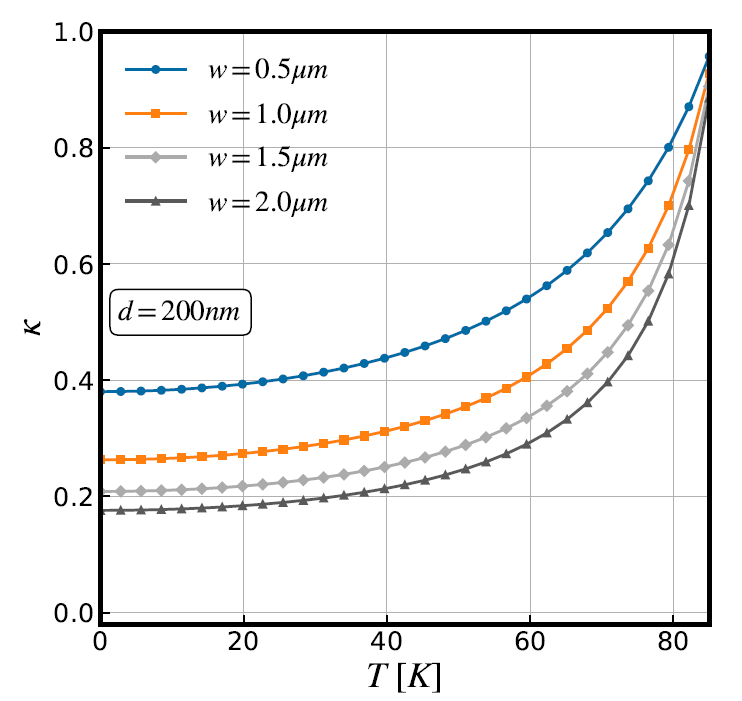}
    \caption{Normalized contribution of the kinetic inductance as a function of temperature, for devices with different upper track widths $w$ and fixed $l_1=50 \, \upmu$m and film thickness $d=200 \,$nm.}
    \label{fig:kappa_vs_T}
\end{figure}

In Figs.~\ref{fig:kappa_vs_d} and \ref{fig:kappa_vs_T}, we plot the normalized kinetic self-inductance $\kappa = L^k_s/L_s$ where $L^k_s$ is the kinetic self-inductance, Eq.~\eqref{eq:L_ka}, and $L_s$ is the total SQUID self-inductance~\cite{Gali2023b}.
{}
{}
Figures~\ref{fig:kappa_vs_d} and \ref{fig:kappa_vs_T} show that $\kappa$ is larger for designs with thinner tracks $w$ and it decreases increasing the film thickness $d$. 
Thus, $\kappa$ decreases for devices with wide tracks and thick films. 
This means that designs with a smaller kinetic inductance contribution are more robust to temperature fluctuations, since it is the only component of the inductance which is temperature dependent.
On the other hand, to design devices that are temperature sensitive, large $\kappa$ values are needed and thus thinner films are more suitable.

\subsection{Voltage dependence on the SQUID self-inductance}
\begin{figure}[h!]
    \centering
    \includegraphics[width=0.5\textwidth]{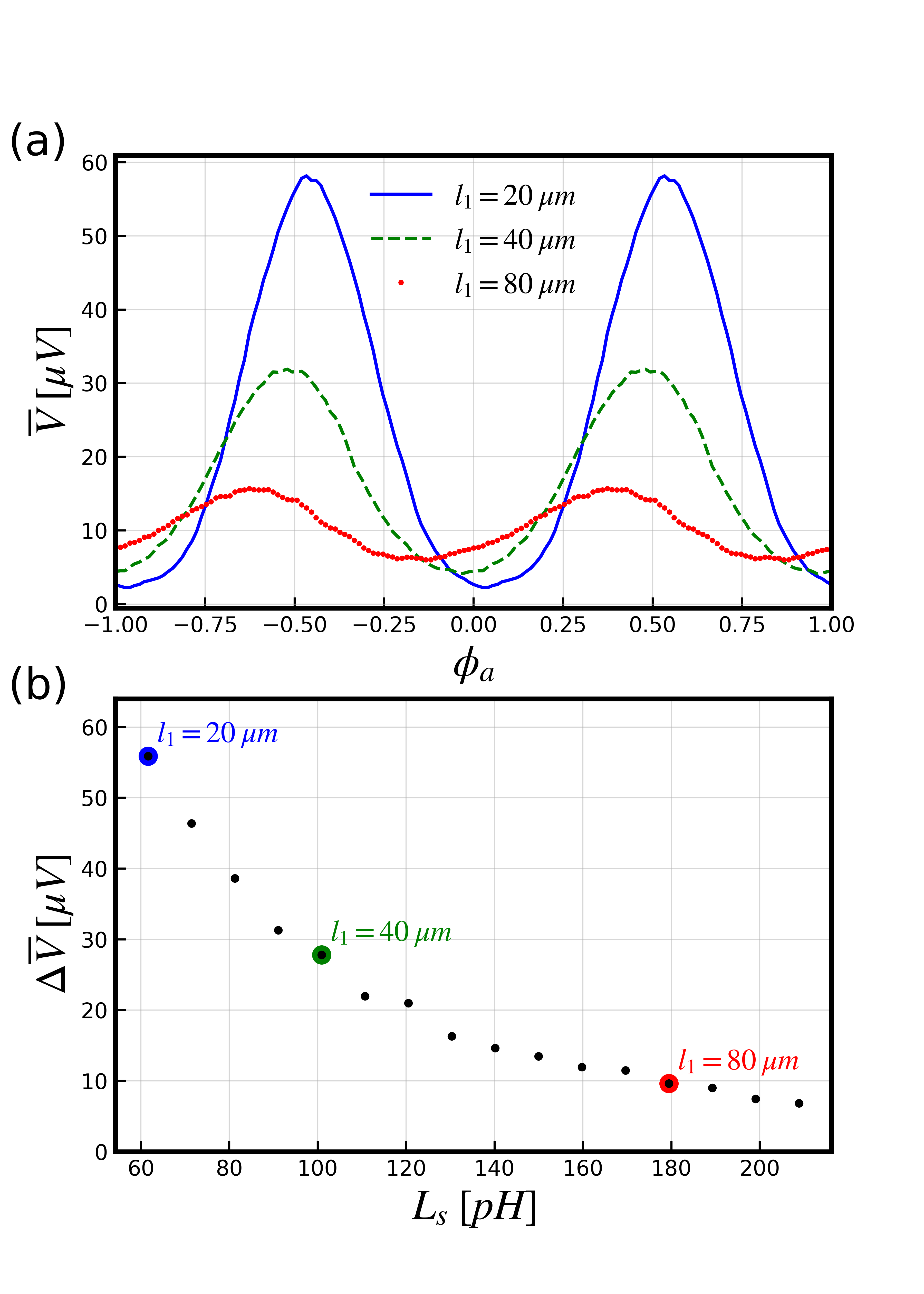}
    \caption{(a) Time-averaged voltage $\overline{V}$ versus normalised applied magnetic flux $\phi_a=\Phi_a/\Phi_0$ of asymmetric SQUIDs with film thickness $d=200 \; $nm and different upper-loop length lengths: $l_1=20\; \upmu $m (blue solid line), $l_1=40\; \upmu $m (green dashed line) and $l_1=80\, \upmu $m (red dotted line). (b) Voltage modulation depth $\Delta \overline{V}$ versus SQUID self-inductance $L_s$. Here $I_c=20 \,\upmu $A, $I_b=40 \, \upmu $A and $T=77 \; $K.}
    \label{fig:V-ind}
\end{figure}

Similarly to the method used in \cite{Gali2022a}, Eq.~\eqref{eq:final_DE} can be solved numerically to obtain the time-averaged voltage response of the SQUID $\overline{V}$. 
Figure \ref{fig:V-ind}(a) shows $\overline{V}$ versus $\phi_a$ of asymmetric SQUIDs operating at $T=77 \, $K and biased at $i_b=0.75$ for three different upper-loop lengths and therefore different asymmetry ratios: $l_1=20 \, \upmu m$ (blue line, $l_1/l_2=2$), $l_1=40 \, \upmu m$ (green dashed line, $l_1/l_2=4$) and $l_1=80 \, \upmu$m (red dotted line, $l_1/l_2=8$). 
These results show that large inductances decrease the device performance, but also that larger $l_1$ produce more fluxoid which shifts the voltage response (clearly seen in the red dotted line in Fig.~\ref{fig:V-ind}(a)).
Figure~\ref{fig:V-ind}(b) shows the dependence of voltage modulation depth on $L_s$ and the colored points shows the devices studied in Fig.~\ref{fig:V-ind}(a). This figure shows more clearly the exponential decay of the device sensitivity with increasing $L_s$, independent of the asymmetry ratio $l_1/l_2$.

{}

\subsection{Voltage modulation depth dependence on the operating temperature}
\begin{figure}[!h]
    \centering
    \includegraphics[width=0.5\textwidth]{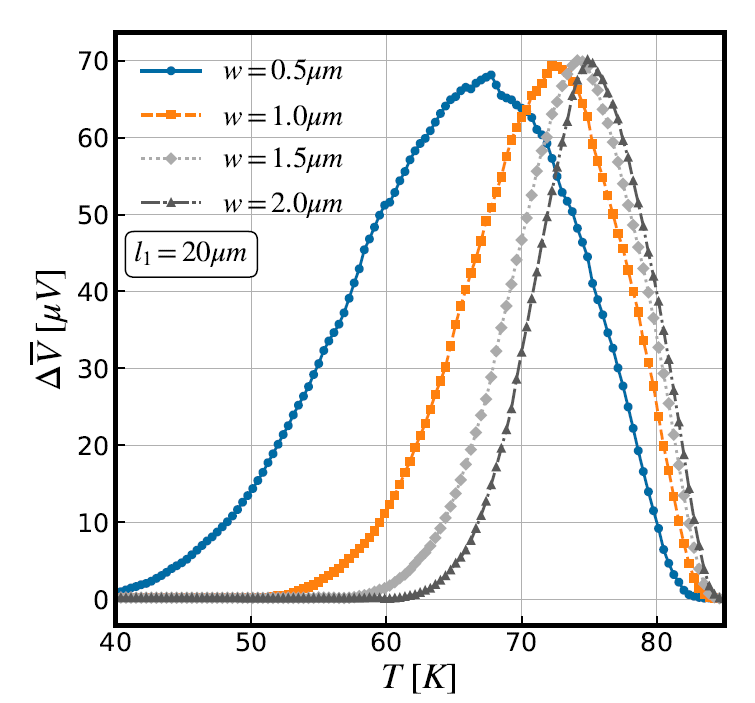}
    \caption{Temperature dependence of the modulation depth $\Delta\overline{V}$ for different track widths $w$ at $i_b=0.75$, using a design with $l_1=20 \, \upmu$m and $d=200 \,$nm.}
    \label{fig:voltage_amplitude_width}
\end{figure}

Using the same approach for generating Fig.~\ref{fig:V-ind}, we can study the effect of changing the temperature and bias current on the device's performance. We see from Fig.~\ref{fig:voltage_amplitude_width} that for a fixed $i_b=0.75$, loop length $l_1$ and film thickness $d$, the voltage modulation depth dependence on $T$, $\Delta\overline{V}(T)$, has a clear maximum which shifts with the upper track width $w$. 
Also, the width of the $\Delta\overline{V}(T)$ curve increases with decreasing $w$.
At $i_b=0.75$, the $\max (\Delta \overline{V})$ values appear when $\beta_L / \Gamma \approx 1$.
This suggests that one can optimise the geometry of the device depending on which temperature we want to operate it. Another approach is to use a SQUID for thermometry. The results in Fig.~\ref{fig:voltage_amplitude_width} show that fixing the inputs $i_b$ and $\phi_a$ we can indirectly estimate the temperature based on $\Delta \overline{V}$.
For instance, the device's temperature sensitivity can be finely-tuned for two different applications: (i) using narrower tracks $\Delta\overline{V}$ presents a broader response in $T$ with a measurable voltage modulation depth in the range $T \in [40,\; T_c)$.
(ii) On the other hand, wider tracks present narrower but steeper $\Delta\overline{V}(T)$, which is suitable when higher device temperature sensitivity is required.

\begin{figure}[!h]
    \centering
    \includegraphics[width=0.5\textwidth]{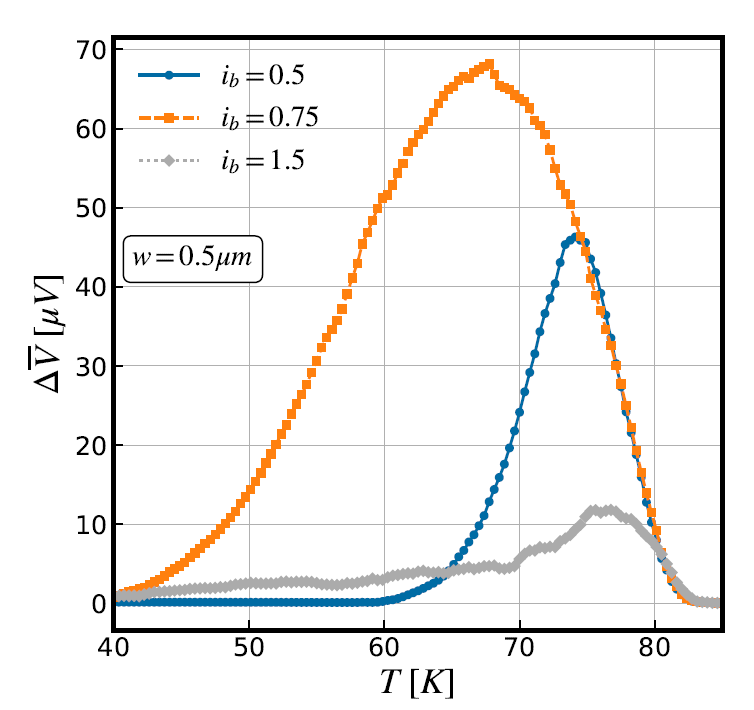}
    \caption{Temperature dependence of the modulation depth $\Delta\overline{V}$ for different $i_b$, using a design with $w=0.5\mu$m, $l_1=20\, \upmu$m and $d=200 \,$nm.}
    \label{fig:voltage_amplitude_ib}
\end{figure}

Figure~\ref{fig:voltage_amplitude_ib} shows $\Delta\overline{V}(T)$ for different $i_b$ values. As reported previously \cite{Gali2022a}, $i_b=0.75$ shows an optimum response when operating close to $T=77 \, $K. Below and above $i_b=0.75$, the response becomes narrower in $T$ and the magnitude of $\Delta\overline{V}$ also decreases substantially, indicating that the choice of $i_b$ plays a crucial role in the device's performance.

\begin{figure}[!h]
    \centering
    \includegraphics[width=0.5\textwidth]{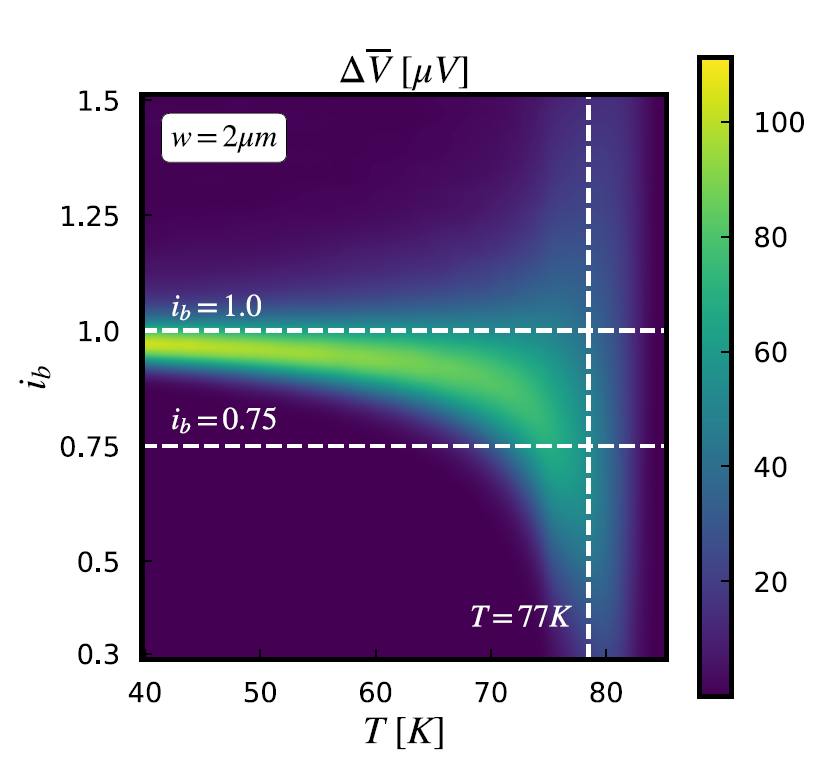}
    \caption{Voltage modulation depth $\Delta\overline{V}$ calculated for different $i_b$ and $T$, using a design with $w=2\mu$m, $l_1=20 \, \upmu$m and $d=200 \,$nm.}
    \label{fig:voltage_modulation_heatmap}
\end{figure}

Figure~\ref{fig:voltage_modulation_heatmap} displays the dependence of $\Delta\overline{V}$ on both $i_b$ and $T$ and shows that the optimal bias current at lower temperatures is $i_b^*\lesssim 1$, while at operating temperatures close to $T_c$ $\Delta\overline{V}$ optimizes at smaller $i_b$. Thus, the optimal bias current depends also on the operating temperature $i_b^*(T)$. 
For instance, at $T=77 \, $K then $i_b^*\approx 0.75$ as shown by the white dashed lines in Fig.~\ref{fig:voltage_modulation_heatmap}, which agrees with previously reported values~\cite{Tesche1977, Gali2022a, Gali2022b}.


\section{Conclusion} \label{sec:Conc}
In this work we have used a mathematical model of an asymmetric SQUID to study the effect of the inductance on the device performance.
By comparing our calculated inductances with experimental measurements, we have shown that the method used is adequate for these kind of structures, showing great accuracy.

From our modelling, as shown in Fig.~\ref{fig:kappa_vs_d}, we found that thinner films produce a stronger dependence on temperature of the kinetic self-inductance ratio $\kappa$, while thicker films have smaller $\kappa$ and therefore are more robust to temperature changes.

We also found that the device response $\overline{V}$ and $\Delta \overline{V}$ with respect to the applied magnetic field is controlled by the total self-inductance, but is unaffected by the device's asymmetry aside from an off-set in the voltage response.

Finally, we analysed the device temperature dependence for different track widths and bias currents.
Our results show that biasing the device at the optimal point is essential for maximising the voltage response, achieving higher sensitivities and broader temperature operation ranges.
We see that the voltage modulation depth shows a bell-like curve, in which the maximum coincides with $\beta_L/\Gamma \approx 1$, when biased at $i_b=0.75$ which is close to the optimal bias current at $T=77\,$K.
These curves indicate that the sensitivity of devices with narrow tracks show a broader temperature response, while wider tracks give a narrower temperature range but higher temperature-sensitivity.
For instance, Fig.~\ref{fig:voltage_amplitude_width} showed that devices with $w=2\, \upmu \,$m present voltage modulation depth variations of tens of micro-volts when temperature changes by only a few Kelvin. On the other hand, devices with $w=0.5 \, \upmu \,$m show a slower variation of $\Delta \overline{V}$ for a wider temperature range.
Our investigation of $\Delta\overline{V}$ depending on $T$ and $i_b$ has shown that at lower temperatures the device voltage modulation depth maximises for $i_b^* \approx 1$, also showing a very large and quick decrease of $\Delta\overline{V}$ for bias current away from the optimal.
This sharp decrease of $\Delta\overline{V}$ with $i_b$ emphasises the importance of correct biasing, which becomes more important as temperature decreases since junction activation due to thermal noise becomes less probable.

\section*{Acknowledgments}
The authors would like to thank K.-H. M\"uller and K. E. Leslie for insightful discussions.

\balance
\bibliography{hairpin_Inductance}

\end{document}